# Dynamics and mechanism of oxygen annealing in $Fe_{1+y}Te_{0.6}Se_{0.4}$ single crystal

Yue Sun[1,2], Yuji Tsuchiya[1], Toshihiro Taen[1], Tatsuhiro Yamada[1], Sunseng Pyon[1], Akira Sugimoto[3], Toshikazu Ekino[3], Zhixiang Shi[2*], Tsuyoshi Tamegai[1†]

[1]Department of Applied Physics, The University of Tokyo, 7-3-1 Hongo, Bunkyo-ku, Tokyo 113-8656, Japan

[2]Department of Physics and Key Laboratory of MEMS of the Ministry of Education, Southeast University, Nanjing 211189, People's Republic of China

[3]Graduate School of Integrated Arts and Sciences, Hiroshima University, Higashi-Hiroshima 739-8521, Japan

Email addresses: *zxshi@seu.edu.cn  †tamegai@ap.t.u-tokyo.ac.jp

## *Abstract*

Iron chalcogenide Fe(Te,Se) attracted much attention due to its simple structure, which is favorable for probing the superconducting mechanism. Its less toxic nature compared with iron arsenides is also advantageous for applications of iron-based superconductors. By intercalating spacer layers, superconducting transition temperature has been raised over 40 K. On the other hand, the presence of excess Fe is almost unavoidable in Fe(Te,Se) single crystals, which hinders the appearance of bulk superconductivity and causes strong controversies over its fundamental properties. Here we report a systematical study of $O_2$-annealing dynamics in $Fe_{1+y}Te_{1-x}Se_x$ by controlling the amount of $O_2$, annealing temperature, and time. Bulk superconductivity can be gradually induced by increasing the amount of $O_2$ and annealing time at suitable temperatures. The optimally annealed crystals can be easily obtained by annealing with ~ 1.5% molar ratio of oxygen at 400 °C for more than 1 hour. Superconductivity was witnessed to evolve mainly from the edge of the crystal to the central part. After the optimal annealing, the complete removal of excess Fe was demonstrated via STM measurements. Some fundamental properties were recharacterized and compared with those of as-grown crystals to discuss the influence of excess Fe.

Superconductivity discovered in iron chalcogenides has stimulated great interests since it is a possible candidate to break the superconducting transition temperature record ($T_c \sim 55$ K) in the iron-based superconductors. Although the initial $T_c$ in FeSe is just 8 K, it increased up to 14 K[1,2] with appropriate Te substitution and 37 K[3,4] under high pressure. Furthermore, by intercalating spacer layers between adjacent FeSe layers, $T_c$ has reached $\sim 32$ K[5] in $A_x$Fe$_{2-y}$Se$_2$ ($A$=K, Cs, Rb and Tl) and 43 K[6] in Li$_x$(NH$_2$)$_y$(NH$_3$)$_{1-y}$Fe$_2$Se$_2$ ($x \sim 0.6$; $y \sim 0.2$). By applying pressure to $A_x$Fe$_{2-y}$Se$_2$, $T_c$ can even reach $\sim 48$ K[7]. Recent angle-resolved photoemission spectroscopy (ARPES) revealed unexpected large superconducting gap $\sim 19$ meV in a single-layer FeSe, which suggests a $T_c$ as high as 65 K[8]. Among iron chalcogenides, Fe$_{1+y}$Te$_{1-x}$Se$_x$ are unique in their structural simplicity, composing of only iron-chalcogenide layers, which is favorable for probing the mechanism of superconductivity. Band structure calculation[9,10] and ARPES[11-13] results show that Fermi surface of FeTe$_{1-x}$Se$_x$ is characterized by hole bands around $\Gamma$ point and electron bands around $M$ point, which is similar to iron pnictides. Moreover, the competition between magnetism and superconductivity, similar to iron pnictides, is also observed in iron chalcogenides[14]. The parent compound Fe$_{1+y}$Te is not superconducting, but exhibits a spin-density wave (SDW) ground state. With Se doping, superconductivity emerges and $T_c$ goes up to 14 K accompanied by the suppression of SDW. Iron chalcogenides also manifest some differences compared with iron pnictides. In particular, Fe$_{1+y}$Te$_{1-x}$Se$_x$ exhibits a bi-collinear antiferromagnetic ordering and it can be either commensurate or incommensurate depending on the Fe content[15,16], different from the common collinear commensurate antiferromagnetic ordering observed in all iron pnictides.

Although much attention has been paid to this compound, some crucially fundamental physical properties are still controversial. In optimally-doped Fe$_{1+y}$Te$_{1-x}$Se$_x$, ARPES[13] shows an isotropic superconducting gap, while anisotropic or two-gap features were suggested by angle-resolved specific heat measurement[17], muon spin rotation measurements[18,19], and optical conductivity measurements[19,20]. Recently, London penetration depth measurement down to 50 mK indicates a possible nodal gap[21]. On the other hand, although Liu *et al.*[14] reported that bulk superconductivity resides only in the region of Se level higher than 29%, it was observed in the Se doping level between 10 and 50% by Noji *et al.*[22,23]. As for the Hall coefficient, conflicting low-temperature behaviors have been observed in crystals with nominally the same amount of Se[24-26]. Even in the case of resistivity, both the metallic and nonmetallic behaviors were reported, and the absolute value just above $T_c$ has a spread from 200 to 1500 $\mu\Omega$ cm[1,2,23-26]. These controversies are believed to come from the sample-dependent Fe nonstoichiometries,[15,27] which originate from the partial occupation of the second Fe site (excess Fe site) in the Te/Se layer. The excess Fe with valence near $Fe^+$ will provide electron doping into the 11 system. The excess Fe is also strongly magnetic, which provides local moments that interact with the adjacent Fe layers[28]. The magnetic moment from excess Fe will act as a pair breaker and also localize the charge carriers[24]. Thus the existence of excess Fe complicates the study of iron chalcogenides from the mechanism of superconductivity to normal state properties. Unfortunately, the excess Fe is easily formed in the standard growth technique employing slow cooling[29]. Specifically, crystals with higher Te content usually contain more excess Fe, and the magnitude of excess Fe decreases with the increase in Se doping[29]. The excess Fe may be also responsible for the unexpected lagged application research despite its simpler structure, less toxic nature, and easier synthesis technology compared with iron-based 122 compounds[30,31]. Thus, eliminating the influence of excess Fe is essential to clarify the intrinsic mechanism of superconductivity as well as to implement applications of iron chalcogenides.

Some previous studies indicated that annealing in $O_2$[32-34], $I_2$[35] and chalcogen[36,37] atmosphere is effective to induce superconductivity in the as-grown $Fe_{1+y}Te_{1-x}Se_x$ single crystal. In particular, our previous studies[34] have demonstrated that proper $O_2$-annealing is very effective in inducing bulk superconductivity in $Fe_{1+y}Te_{0.6}Se_{0.4}$ single crystal. In this report, we have systematically studied the optimum condition and dynamics of $O_2$-annealing to induce bulk superconductivity. $O_2$-annealing is proved to be effective to remove the excess Fe, and a pure crystal without excess Fe was obtained. In addition to the evolution of superconductivity under $O_2$-annealing, changes in some fundamental properties after removing the excess Fe were also studied and presented in the paper.

## Results

**Optimum condition and dynamics of $O_2$-annealing in $Fe_{1+y}Te_{0.6}Se_{0.4}$** To study the optimum condition and dynamics of $O_2$-annealing, we firstly fixed the annealing temperature at 400 °C, and annealed the crystals by successively increasing amount of $O_2$ for 20 hours (enough time for the crystals to absorb the $O_2$ totally). Shown in Figure 1a is the temperature dependence of zero-field-cooled (ZFC) and field-cooled (FC) magnetization at 5 Oe for $Fe_{1+y}Te_{0.6}Se_{0.4}$ single crystal annealed with increasing the amount of $O_2$. The as-grown crystal usually shows no superconductivity or very weak diamagnetic signal below ~ 3 K. After annealing, superconductivity emerges and $T_c$ is gradually enhanced with increasing the cumulative amount of $O_2$. The diamagnetic signal also increases with increasing the amount of $O_2$ even faster than the increase of $T_c$. Its magnitude at low temperatures reaches the maximum value when the onset $T_c$ reaches ~ 7 K. However, it is well known that even the sample is mostly non-superconducting, the diamagnetic signal becomes significant when the non-superconducting region is covered by superconducting region. Thus, to prove the bulk superconductivity induced by $O_2$-annealing, we need to refer to other quantities. Although specific heat is ideal for this purpose, limited size of the crystal optimized for uniform annealing does not allow reliable evaluation of the volume fraction of superconductivity. As an excellent alternative to probe bulk superconductivity, we propose to use the critical current density $J_c$. The method used for the calculation of $J_c$ can be seen in the supplement. Self-field $J_c$ at 2 K vs. the amount of $O_2$ are summarized in Figure 1b together with the change in $T_c$. It is obvious that both $T_c$ and $J_c$ are gradually enhanced by annealing with increasing amount of $O_2$, and reach the opt FeAs layers imal values of $T_c$ higher than 14 K and $J_c$ larger than $1\times10^5$ A/cm$^2$ when the ratio of oxygen to the total Fe is ~ 1.3% to 1.5%. Bulk superconductivity with high $T_c$ and large $J_c$ is induced by the removal of excess Fe as we will show later. Here we should point out that when the ratio of oxygen is higher than 1.3%, the value of $J_c$ slightly decreases, which may come from the reaction of Fe in Fe layers with $O_2$. Actually, we found the crystal will be decomposed into oxides such as $Fe_2O_3$, $TeO_2$ and $SeO_2$ after annealing with too much $O_2$, as shown in the X-ray diffraction (XRD) pattern of Figure S1.

Next we study the effect of annealing temperature and time. In this experiment, the crystals were annealed with the optimal amount of $O_2$ (ratio of oxygen to the nominal Fe is 1.5%) at different temperatures with increasing time. It should be noted that the total amount of $O_2$ reacted with the crystal is larger than 1.5% since we fix the amount of $O_2$ in all the annealing processes. Shown in Figure 2a is the typical result of annealing at 200 °C. With increasing annealing time, $T_c$ gradually shifts to higher temperatures together with the sharpening of the transition width. $J_c$ is also measured to probe the evolution of bulk superconductivity, which is shown in Figure 2b together with the change in $T_c$. Both $T_c$ and $J_c$ are increased with increasing annealing time, and reach the optimal values after 10 and 100 hours, respectively. To know the influence of temperature, the crystals were also annealed at 250, 300 and 400 °C with increasing time, and one more typical results of a crystal annealed at 400 °C are shown in Figure 2c. At all these annealing temperatures, both $T_c$ and $J_c$ are gradually increased with annealing time similar to the case of 200 °C. Changes in $J_c$ with annealing time at different temperatures were further scaled by $J_c/J_c(max)$, and plotted in Figure 2d, which manifests that the rate of inducing superconductivity is accelerated at higher annealing temperatures. It also proves that $O_2$-annealing is very effective to induce bulk superconductivity in $Fe_{1+y}Te_{0.6}Se_{0.4}$. In particular, in the case of 400 °C, $T_c$ and $J_c$ reach the optimal values only after 5 and 60 min annealing, respectively. We also tried to anneal the crystal at higher temperatures, like 500 °C. In this case, the crystal is damaged during the annealing and superconductivity is not induced.

**Spatial evolution of superconductivity during annealing** The above results manifest that $O_2$-annealing can gradually induce bulk superconductivity in $Fe_{1+y}Te_{0.6}Se_{0.4}$, while the spatial evolution of superconductivity during the annealing and distribution of $J_c$ are still unknown, which is crucial to the understand of the mechanism of $O_2$-annealing. In order to reveal these unresolved issues, we employ the magneto-optical (MO) imaging for a crystal annealed for different periods of time, which allows direct observations of the spatial evolution of superconductivity and $J_c$. Figure 3a and b show the MO images of flux expulsion under 5 Oe at 10 K for the crystal annealed at 400 °C for 1 and 3min, respectively. When the sample is annealed for 1 min, only the edge part shows Meissner state, while the central part still remains non-superconducting. When the annealing time is increased to 3 min, all parts of the sample turn into Meissner state. It indicates that the superconductivity mainly evolves from edges of the crystal to the center.

As we know from Figure 2c, $J_c$ of the crystal annealed at 400 $^o$C for 3 min is still lower than the optimal value, which means if the superconducting state evolves from the edge to the center, the edge of the crystal should maintain larger $J_c$ than the central region. To examine this assumption, we also took MO images on the crystal annealed for 3 min in the remanent state at temperatures ranging from 5 to 14 K, which are shown in Figure S2 in the supplement. Figure 3c shows profiles of the magnetic induction at different temperatures along the dashed line shown in Figure S2a. Both the MO images and profiles show that the magnetic field cannot totally penetrate the sample at 5 K. At higher temperature, the field penetrates deeper into the center. At 8 K, the sample is totally penetrated, and the MO image manifests a typical roof-top pattern up to 10 K. After that, from 11 to 14 K, the edge of the crystal traps higher magnetic field than the center, which indicates $J_c$ in the edge of the crystal is larger. To directly observe the distribution of $J_c$ in the crystal, we converted the MO images into the current distribution with a thin-sheet approximation and by using the fast Fourier transform calculation method[38]. A typical image of the distribution of the modulus of $J_c$ at 10 K is shown in Figure 3d, which manifests that the edge of the crystal maintains a larger current, and the $J_c$ decays from the edge to the center. All these results testify that superconductivity mainly evolves from the edge of the crystal to the center. It should be noted, however, that we can observe a total expulsion of magnetic flux at 5 K even in the sample annealed for only 1 min (not shown). This fact means that weak superconductivity also evolves from the top and bottom surfaces.

## Discussion

The above results on the $O_2$-annealing dynamics prove that it is effective to induce bulk superconductivity in $Fe_{1+y}Te_{0.6}Se_{0.4}$ single crystal, and the superconductivity evolves mainly from the edge of the sample to the central part. Now we turn to the discussion of the mechanism of $O_2$-annealing. Single crystal XRD measurements show that lattice constant $c$ keeps almost constant value during the annealing process as shown in Figure S3a. On the other hand, the lattice constant $a$ increases from 3.802 ± 0.002 Å for the as-grown crystal to 3.812 ± 0.002 Å for the annealed one. Such an increase in the in-plane lattice constant after removing excess Fe is similar to previous reports on structure refinement of crystals containing different amount of excess Fe[35,39]. However, it is different from the structure change of $FeTe_{0.8}S_{0.2}$ annealed in $O_2$[40]. Actually, in $FeTe_{0.8}S_{0.2}$, the superconductivity induced by $O_2$-annealing is reported to be removed by further vacuum annealing[40]. We also annealed our $O_2$-annealed crystal again in vacuum at 400 °C for more than two weeks. The magnetization results for a crystal before and after vacuum annealing were compared in Figure S3b, which shows that the superconductivity induced by $O_2$-annealing is stable under further vacuum annealing. This result together with the unchanged lattice constant $c$ all indicates that the oxygen is not simply doped into the sample.

As is well known, the presence of excess Fe is unavoidable in the growth of $Fe_{1+y}Te_{1-x}Se_x$ single crystals, which suppresses the superconductivity and causes magnetic correlations[27]. In our as-grown single crystals, the amount of excess Fe is ~ 14% as analyzed by Inductively-coupled plasma (ICP) atomic emission spectroscopy. The $O_2$-annealing may induce superconductivity by removing excess Fe from the as-grown crystal. Although the excess Fe may be removed after $O_2$-annealing, it should still remain in the crystal, mainly on the surface, in some form of oxides. Thus, traditional compositional analysis methods like ICP, energy dispersive X-ray spectroscopy (EDX) and electron probe microanalyzer (EPMA) can hardly detect the change of excess Fe. Actually our ICP analyses on the $O_2$-annealed crystal just show a slight change in the Fe content compared with the as-grown one. One important information is the fact that the color of the surface layer of the crystal gradually changes during the $O_2$-annealing and turns into blue after annealing with sufficient $O_2$ as shown in the Figure S4, which is similar to the previous report on the crystal annealed in air[41]. EDX results show that the blue surface contains more Fe, which is interpreted as a signature that removed excess Fe forms oxide on the surface after annealing. To precisely determine the change in the number of excess Fe, we employ the scanning tunneling microscopy (STM) measurement, which has atomic resolution. The excess Fe occupies the interstitial site in the Te/Se layer[42], and the previous report proved that the cleaved $Fe_{1+y}Te_{1-x}Se_x$ single crystal possesses only the termination layer of Te/Se[43], which guarantee the STM can directly observe the excess Fe in Te/Se layer without the influence of neighboring Fe layers. Shown in Figure 4a is the STM image for the as-grown $Fe_{1+y}Te_{0.6}Se_{0.4}$ single crystal. There are several bright spots in the image, which represent the excess Fe according to the previous STM analysis[43-47]. After optimally annealing in $O_2$, the bright spots, i.e. the excess Fe, disappeared in STM images as shown in Figure 4b and a larger region in Figure 4c. Actually, we searched several different regions, and almost no bright spots can be found, which directly proves that the excess Fe is almost totally removed by $O_2$-annealing. It may be also the reason why our $O_2$-annealed $Fe_{1+y}Te_{0.6}Se_{0.4}$ single crystal is of high quality.

Figure 5a shows the temperature dependence of in-plane resistivity for the as-grown and $O_2$-annealed $Fe_{1+y}Te_{0.6}Se_{0.4}$ single crystals. For both as-grown and $O_2$-annealed crystals, resistivity maintains a nearly constant value above ~ 150 K. On the other hand, from 150 K down to the superconducting transition, the as-grown crystal shows a nonmetallic behavior ($d\rho/dT < 0$), which is replaced by a metallic behavior ($d\rho/dT > 0$) after removing the excess Fe by $O_2$-annealing. Similar results are also reported in $Fe_{1+y}Te_{1-x}Se_x$ with different amounts of excess $Fe^{[24,48]}$. To probe the influence of excess Fe to the superconductivity, specific heat is carefully measured in both the as-grown and $O_2$-annealed $Fe_{1+y}Te_{0.6}Se_{0.4}$ single crystals. Specific heat divided by temperature $C/T$ as a function of $T$ for both crystals are shown in Figure 5b. Obviously, no specific heat jump can be observed in the as-grown crystal, which proves that the superconducting transition observed in the resistivity measurement comes from the filamentary superconductivity. After $O_2$-annealing, a clear specific heat jump can be observed, indicating the bulk nature of the superconductivity in $O_2$-annealed $Fe_{1+y}Te_{0.6}Se_{0.4}$. Specific heat jump $\Delta C/T_c$ was obtained as ~ 66.3 mJ/molK$^2$ with $T_c$ = 14.5 K determined by entropy balance consideration. The normalized specific heat jump at $T_c$, $\Delta C/\gamma_n T_c$, is ~ 3.0 by using the fitted Sommerfeld coefficient of $\gamma_n$ = 22 mJ/molK$^2$. (Details about the fitting of the specific heat data and more discussions can be seen in the supplement.) The value is larger than those of 2.1 ~ 2.8 in previous reports[22,49], and is much larger than BCS weak-coupling value of 1.43 indicating that superconductivity in $O_2$-annealed $Fe_{1+y}Te_{0.6}Se_{0.4}$ is strong coupling in nature. To further reveal the influence of excess Fe on the electronic behavior, Hall coefficient $R_H$ of the as-grown and well-annealed samples were measured and shown in Figure 5c. $R_H$ is almost temperature independent above 150 K, and keeps a constant value $\sim 1\times10^{-9}$ m$^3$/C for both the as-grown and O$_2$-annealed samples. When temperature decreases below 150 K, an obvious divergence in $R_H$ is observed. For the as-grown crystal, $R_H$ gradually increases with decreasing temperature. On the other hand, for the O$_2$-annealed crystal, $R_H$ keeps nearly temperature independence above 50 K, followed by a sudden decrease, and finally changes sign from positive to negative before approaching $T_c$. The sign reversal in Hall coefficient can be attributed to the multiband structure[12], indicating the dominance of electron in the charge carriers before the occurrence of superconductivity. We should point out that the sign reversal in $R_H$ is also observed in high-quality polycrystal and thin films[26,50], usually absent in single crystals with more excess Fe[24], which may be the intrinsic nature of FeTe$_{0.6}$Se$_{0.4}$. Figure 5d shows the comparison of high-temperature magnetic susceptibility, $\chi$, for Fe$_{1+y}$Te$_{0.6}$Se$_{0.4}$ single crystals before and after the O$_2$-annealing. At high temperatures, susceptibilities for both crystals maintain almost constant values, and the value of $\chi$ in the O$_2$-annealed crystal is appreciably larger than that of the as-grown crystal, distinct from the previous report[51], which may come from the contribution of iron oxide on the surface. When the temperature approaches $T_c$, $\chi$ slightly decreases in the annealed crystal, while it dramatically increases in the as-grown sample. The differences in resistivity, specific heat, Hall effect, and magnetic susceptibility in Fe$_{1+y}$Te$_{0.6}$Se$_{0.4}$ before and after the O$_2$-annealing can be attributed to the delocalization of excess Fe by the O$_2$-annealing. Density functional study shows that the excess Fe in the interstitial sites is magnetic and interacts with magnetism of Fe in Fe planes[28]. The magnetic moment from excess Fe will act as a pair breaker, so that the specific heat jump associated with the superconducting transition is suppressed in the as-grown crystal and recurs after removing the excess Fe by O$_2$-annealing. On the other hand, the magnetic moment will also localize the charge carriers, which causes the nonmetallic resistive behavior and the increase of $R_H$ at low temperatures. Moreover, neutron scattering measurements revealed that the coupling between the excess Fe and adjacent Fe layers will tune the AFM order from commensurate to incommensurate[15,16]. Furthermore, the resistivity, Hall coefficient, and magnetic susceptibility for the as-grown crystal all undergo a steep increase below 50 K, which is consistent with the neutron scattering result that the short-range magnetic order in a crystal with rich excess Fe enhances noticeably below 40 K[15,52]. On the other hand, the divergence of transport properties between the as-grown and $O_2$-annealed samples occurs below 150 K, which may come from the increasing correlation below that temperature[53].

In summary, we have systematically studied the optimal condition and dynamics of $O_2$-annealing in $Fe_{1+y}Te_{0.6}Se_{0.4}$ single crystal. Bulk superconductivity was gradually induced by annealing with increasing amount of $O_2$ and annealing time at temperatures from 200 ~ 400 °C. MO images show that superconductivity mainly evolves from the edge of the crystal to the center. STM observations directly demonstrate that the excess Fe is removed after the optimal annealing. Transport, heat capacity, and magnetization measurements on the sample before and after the $O_2$-annealing manifest that the $O_2$-annealing induces bulk superconductivity and delocalize the charge carriers by removing the excess Fe. The controllable way of obtaining high-quality Fe(Te,Se) crystals without excess Fe is hopeful to solve the controversies in iron chalcogenides, and to pave the way to overcome the obstacles for applications.

## Methods

**Sample growth and $O_2$-annealing.** Single crystals with a composition $Fe_{1+y}Te_{0.6}Se_{0.4}$ were grown by using the self-flux method as described in detail elsewhere[34]. The obtained crystals were then cut and cleaved into thin slices with dimensions about $2.0 \times 1.0 \times 0.05$ mm$^3$, weighed and loaded into a quartz tube with $d \sim 10$ mm $\phi$ for further annealing. The quartz tube was carefully baked and tested not to emit any gas under the same condition as the sample annealing. The quartz tube was carefully evacuated by a diffusion pump, and the pressure was measured by a diaphragm-type manometer with an accuracy better than 1 mTorr. After totally pumping out the gas, we filled $Ar/O_2(1\%)$ mixed gas into the quartz tube and sealed it into the length of 100 mm. During the sealing process, the manometer monitored the pressure in the system in real-time to prevent gas leakage and controlled the $O_2$ pressure in the quartz tube. Because of the fixed volume, the amount of $O_2$ filled into the quartz tube can be evaluated by the pressure. Then the crystals were annealed at selected temperatures (ranging from 200 °C to 400 °C) for different periods of time, followed by water quenching. We also confirmed that almost all $O_2$ in the quartz tube were consumed by the crystal by breaking the quartz tube in a larger quartz tube while monitoring its pressure. In this case, we used pure $O_2$ gas. Furthermore, we confirmed that the increase in the crystal weight after the annealing is equal to the mass of $O_2$ filled into the quartz tube. Our careful experiments described above guarantee that the $O_2$ was totally absorbed by the crystal.

**Measurements and verifications.** Magnetization measurements were performed to check the superconducting transition temperature $T_c$ and critical current density $J_c$ by using a commercial superconducting quantum interference device (SQUID). Measurements on sample annealed at certain temperature with increasing amount of $O_2$, or increasing annealing time were performed on a given crystal piece to avoid piece-dependent influence. Longitudinal and transverse (Hall) resistivity measurements were performed by the six-lead method. Specific heat was measured with a Quantum Design physical property measurement system (PPMS). Magneto-optical images were taken by using the local-field-dependent Faraday effect in the in-plane magnetized garnet indicator film employing a differential method. Details of the lattice constant change during the annealing was characterized by means of X-ray diffraction (MAC Science M18XHF) with Cu-$K\alpha$ radiation. Inductively-coupled plasma (ICP) atomic emission spectroscopy and energy dispersive X-ray spectroscopy (EDX) were used for chemical analyses. Scanning tunneling microscopy (STM) images were obtained by a modified Omicron LT–UHV–STM system[54]. The sample was cleaved in situ at 4 K in an ultra-high vacuum chamber of ~ $10^{-8}$ Pa to obtain fresh and unaffected crystal surface.

## Acknowledgements

This work is partly supported by the Natural Science Foundation of China, the Ministry of Science and Technology of China (973 project: No. 2011CBA00105), and Japan-China Bilateral Joint Research Project by the Japan Society for the Promotion of Science. ICP analyses were performed at Chemical Analysis Section in Materials Analysis Station of NIMS.

## Author contributions

Y.S performed most of the experiments and analyzed the data. Y.T contributed to the MO images. T. Taen contributed to the growth of single crystal. T.Y and S.P contributed to the specific heat measurement. A.S and T.E performed the STM measurement. Y.S, Z.S and T. Tamegai designed the research. Most of the text of the paper was written jointly by Y.S and T. Tamegai. All the authors contributed to discussion on the results for the manuscript.

## Author Information

The authors declare that they have no competing financial interests. Correspondence and requests for materials should be addressed to Z.S (zxshi@seu.edu.cn) or T. Tamegai (tamegai@ap.t.u-tokyo.ac.jp)

# Figure captions

Figure 1: (a) Temperature dependence of zero-field-cooled (ZFC) and field-cooled (FC) magnetization at 5 Oe for $Fe_{1+y}Te_{0.6}Se_{0.4}$ single crystal annealed at 400 °C with increasing amount of $O_2$ (molar ratio of the oxygen to the nominal Fe ranging from 0.1% to 1.5%). (b) Changes in $T_c$ and self-field $J_c$(2 K) for $Fe_{1+y}Te_{0.6}Se_{0.4}$ annealed at 400 °C with increasing amount of $O_2$.

Figure 2: (a) Temperature dependence of magnetization at 5 Oe for $Fe_{1+y}Te_{0.6}Se_{0.4}$ single crystal annealed at 200 °C with increasing time. Changes in $T_c$ and self-field $J_c$(2 K) with increasing annealing time at (b) 200 °C, (c) 400 °C. (d) Changes in $J_c^{2K}(0)/J_{c,\,max}^{2K}(0)$ with increasing the annealing time at temperature from 200 °C to 400 °C.

Figure 3: Meissner state magneto-optical (MO) images under 5 Oe at 10 K for $Fe_{1+y}Te_{0.6}Se_{0.4}$ annealed at 400 °C for (a) 1 and (b) 3min, respectively. (c) Local magnetic induction profiles at temperatures from 5 K to 14 K taken along the dashed lines in Figure S2a. (d) Spatial distribution of $J_c/J_c$(max) at 10 K for sample annealed at 400 °C for 3 min.

Figure 4: STM images for (a) as-grown, (b) – (c) $O_2$-annealed $Fe_{1+y}Te_{0.6}Se_{0.4}$ single crystal. The bright spots in (a) correspond to the excess Fe, which disappear in the optimally-annealed crystal.

Figure 5: Temperature dependence of (a) in-plane resistivity scaled by the value at 300 K, (b) specific heat plotted as $C/T$ vs $T$, (c) Hall coefficient, and (d) magnetization under 50 kOe with field parallel to $c$-axis in the as-grown and $O_2$-annealed $Fe_{1+y}Te_{0.6}Se_{0.4}$ single crystals.

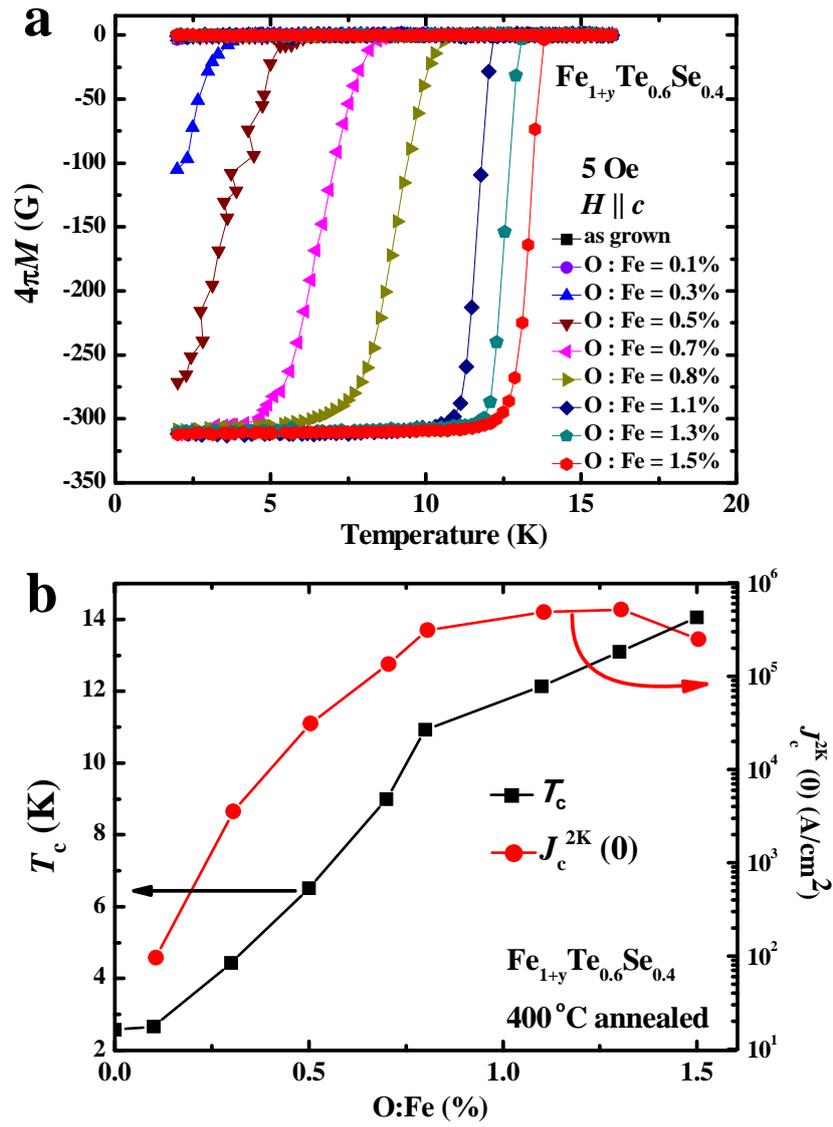

**Figure 1**

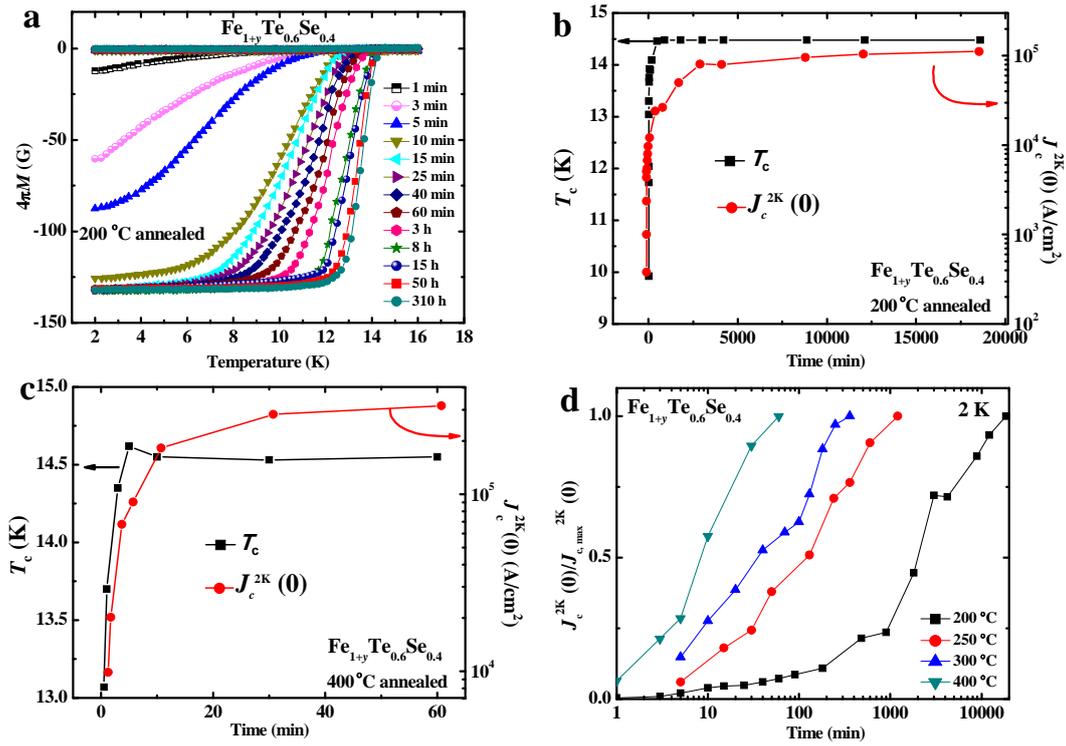

**Figure 2**

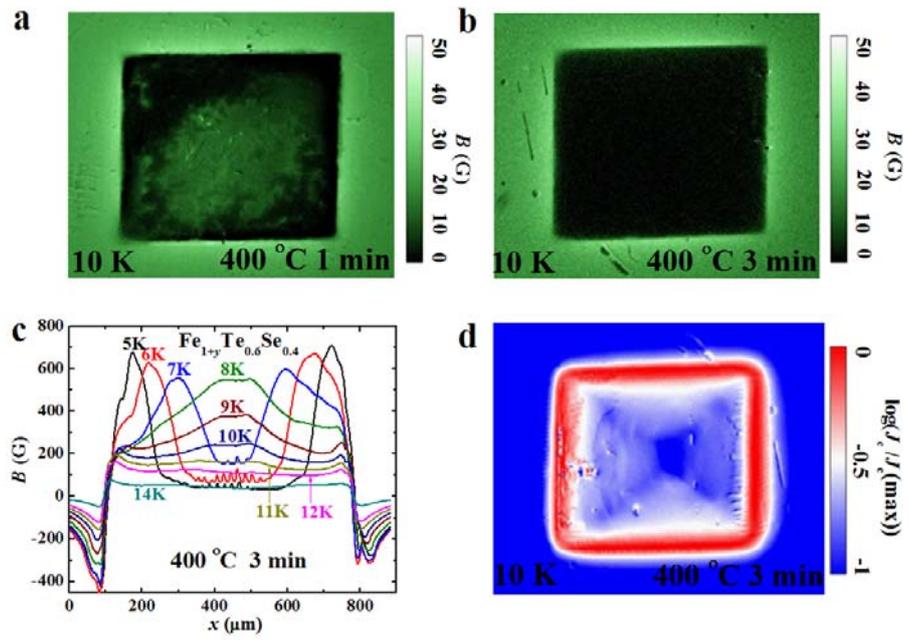

**Figure 3**

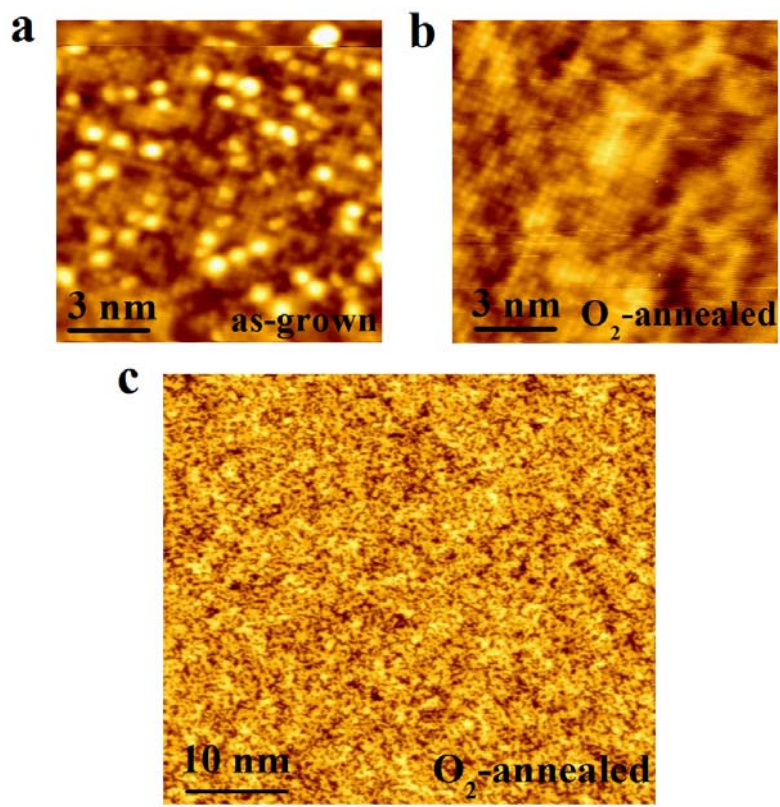

**Figure 4**

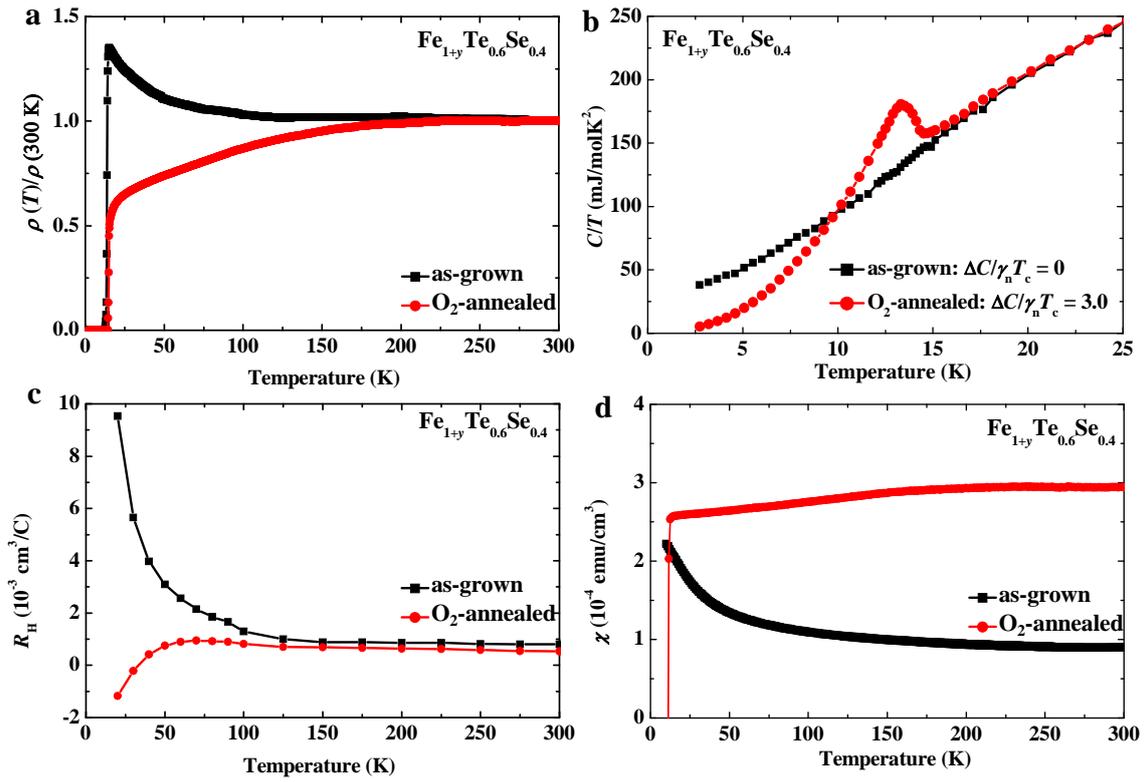

**Figure 5**

# Supplementary Information

## 1. Calculation of $J_c$

$J_c$ [A/cm$^2$] can be evaluated from magnetic hysteresis loops (MHLs) using the Bean model[1]:

$$J_c = 20 \frac{\Delta M}{a(1-a/3b)},$$

where $\Delta M$ [emu/cm$^3$] is $M_{down} - M_{up}$, $M_{up}$ [emu/cm$^3$] and $M_{down}$ [emu/cm$^3$] are the magnetization when sweeping fields up and down, respectively, $a$ [cm] and $b$ [cm] are sample widths ($a < b$).

**2. XRD pattern of the crystal over-annealed with too much O$_2$.**

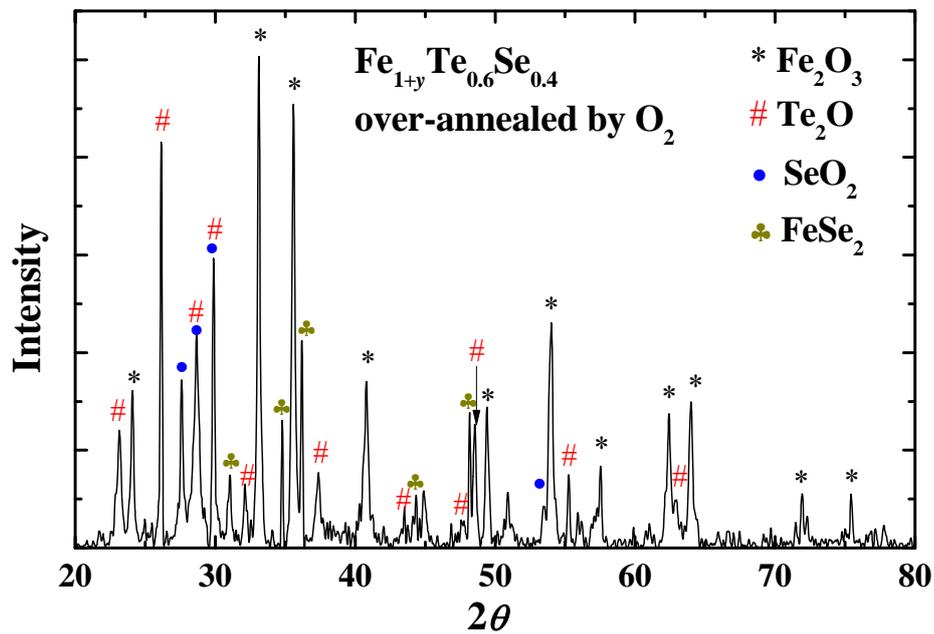

Figure S1 : XRD pattern of Fe$_{1+y}$Te$_{0.6}$Se$_{0.4}$ single crystals annealed at 400 °C in 1 atmosphere of pure O$_2$ (continuous supply of O$_2$) for 40 h. Obviously, the crystals were decomposed into oxides such as Fe$_2$O$_3$, TeO$_2$, SeO$_2$ and some amount of FeSe$_2$.

## 3. Magneto-optical images in the remanent state

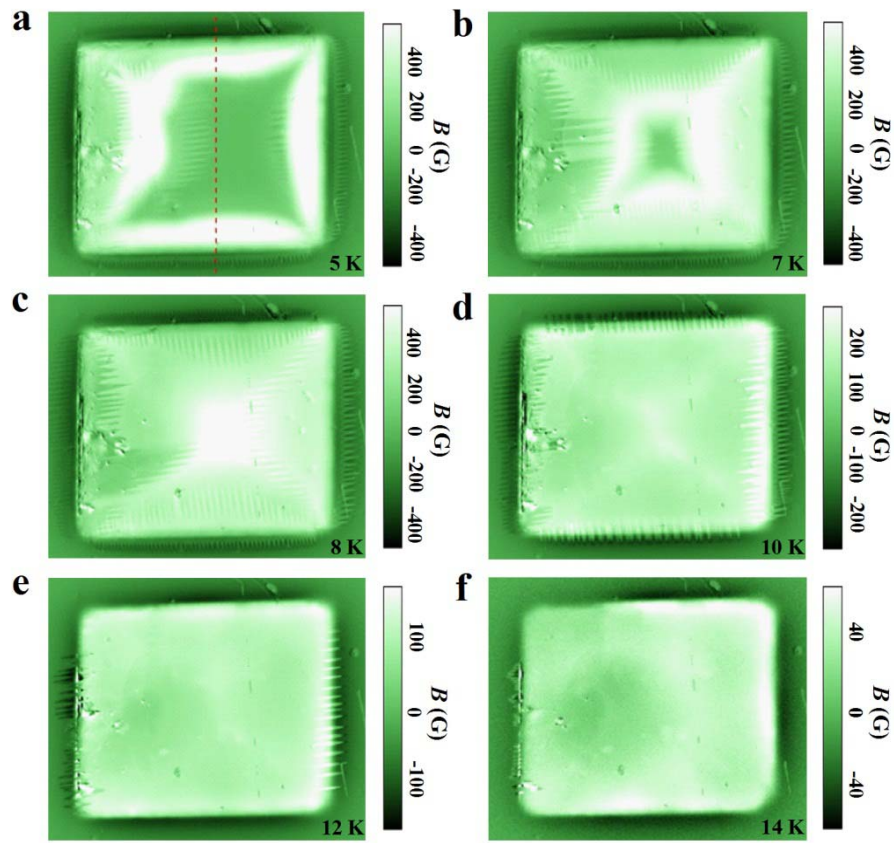

Figure S2 : Magneto-optical images in the remanent state at (a) 5 K, (b) 7 K, (c) 8 K, (d) 10 K, (e) 12 K, and (f) 14 K for $Fe_{1+y}Te_{0.6}Se_{0.4}$ annealed under $O_2$ atmosphere at 400 °C for 3 min. This state is prepared by applying 400 Oe along the *c*-axis for 1 s and removing it after zero-field cooling.

**4. Change in the *c*-axis lattice constant and the effect of vacuum annealing in the annealed crystal**

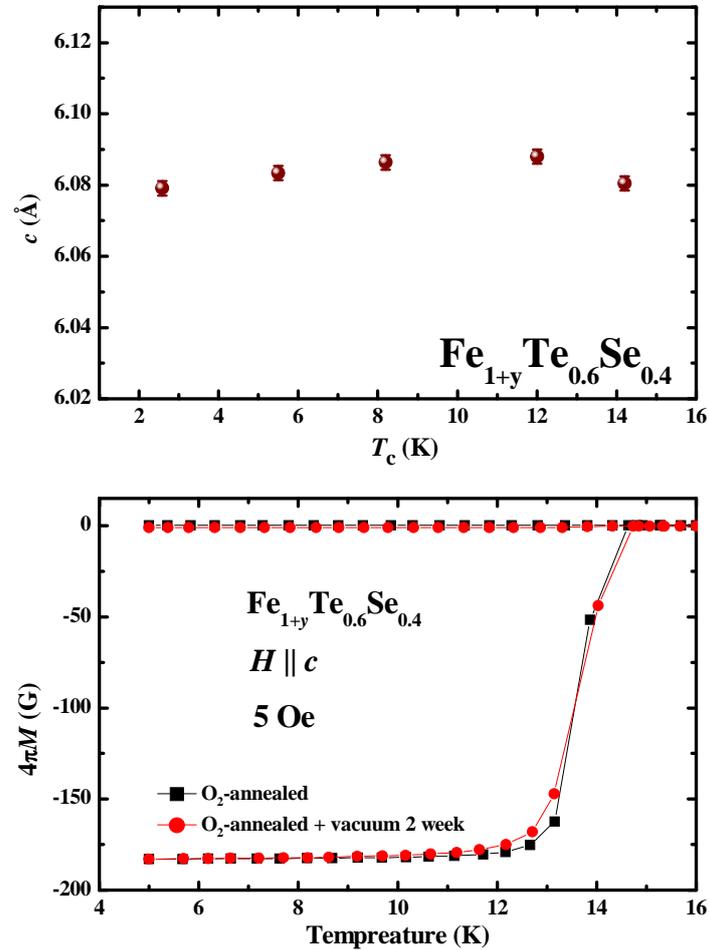

Figure S3: (a) The value of lattice constant $c$ calculated from single crystal XRD on $Fe_{1+y}Te_{0.6}Se_{0.4}$ with different $T_c$'s, which is obtained by annealing with increasing amount of $O_2$. Almost no obvious change of $c$ can be witnessed. (b) Temperature dependence of magnetization at 5 Oe for $O_2$-annealed $Fe_{1+y}Te_{0.6}Se_{0.4}$ single crystal before and after further vacuum annealing.

## 5. Color change of the surface layers during O$_2$-annealing

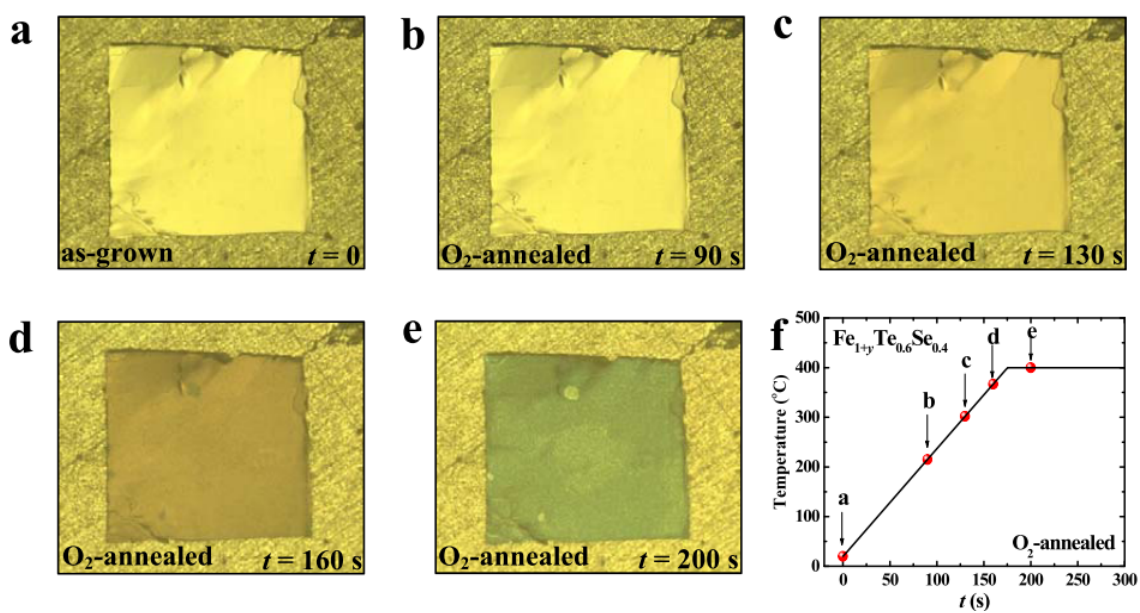

Figure S4: Color change of the surface layer during O$_2$-annealing. The crystal was annealed with enough amount of O$_2$ in a transparent furnace. The furnace was quickly heated up to 400 °C at a rate of 130 °C s$^{-1}$ as shown in (f). During the annealing, a CCD camera was used to acquire the images of the sample surface. Typical images of the crystal before and after annealing for 90 s(~ 200 °C), 130 s(~ 300 °C), 160 s(~ 370 °C), 200 s(~ 400 °C) were show in (a) – (e), respectively. Obviously, the crystal surface first changes color to brown, then turns to blue, which indicates some compounds were formed on the surface during annealing. Here, we should point out that the color change occurs only on the sample surface. After cleaving the surface layers, inner part of the crystal still keeps the mirror-like surface similar to the as-grown crystal.

## 6. Specific heat of $Fe_{1+y}Te_{0.6}Se_{0.4}$ single crystals

Figure S5 shows the temperature dependence of the specific heat plotted as $C/T$ vs $T$ for the as-grown, half- and fully-annealed $Fe_{1+y}Te_{0.6}Se_{0.4}$ single crystals. The half- and fully-annealed crystals were prepared by annealing different pieces of crystals with ~ 0.7% and 1.5% molar ratio of oxygen at 400 °C. In order to avoid the contribution from impurities, surfaces of the half- and full-annealed crystals were removed before the measurement. Obviously, no specific heat jump can be observed in the as-grown crystal, which proves that the superconducting transition observed in the resistivity measurement comes from the filamentary superconductivity. After annealing in $O_2$, a hump-like jump associated with superconducting transition is observed in the half-annealed crystal. (In order to correct the absolute value of the specific heat in the half-annealed crystal due to its relatively small size, we scaled it by multiplying a factor of 1.04.) After further annealing with the optimal amount of $O_2$, a steep superconducting jump can be witnessed in the fully-annealed crystal. The increase of superconducting jump proves that the bulk superconductivity is induced by increasing amount of $O_2$ during the annealing, which is consistent with the conclusion derived from the evolution of $J_c$. The left inset shows the low-temperature specific heat plotted as $C/T$ vs $T^2$ for the three crystals. The residual electronic specific heat coefficient $\gamma_e^{res}$ can be simply obtained by linearly extrapolating the data to zero temperature. (For non-superconducting sample, the obtained value represents the normal state electronic specific heat coefficient $\gamma_{as}$, which is used in the fitting below.) Obviously, $\gamma_e^{res}$ gradually decreases with $O_2$-annealing, and is almost zero for the fully-annealed crystal, manifesting that the superconducting volume reaches ~ 100%. This result also directly proves that the residual electronic specific heat reported before in $Fe_{1+y}Te_{1-x}Se_x$ is coming from the disorders or impurities induced pairing-breaking effect rather than the nodes in superconducting gap, which may be originated from the effect of excess Fe[2-6].

Since the fully-annealed and as-grown crystals share similar phonon specific heat, the phonon contribution for the fully-annealed crystal can be evaluated using the as-grown one as a reference, which can be expressed as $C_{ph}^{fully-annealed}(T) = AC_{ph}^{as-grown}(BT)$, where $A$ and $B$ are the parameters responsible for the sample dependent lattice contribution[4,7,8]. Since the as-grown crystal shows no superconducting jump, the specific heat can be simply fitted by $C = \gamma_{as}T+\beta_n T^3+\alpha_n T^5+\varepsilon_n T^7$ with $\gamma_{as} = 33.5$ mJ/molK$^2$, $\beta_n = 0.69$ mJ/molK$^4$, $\alpha_n = -8.90 \times 10^{-4}$ mJ/molK$^6$ and $\varepsilon_n = 5.06 \times 10^{-7}$ mJ/molK$^8$ between 3 and 25 K. The normal state specific heat of the fully-annealed crystal can be represented by $C_n(T) = C_e^{fully-annealed}(T) + C_{ph}^{fully-annealed}(T) = C_e^{fully-annealed}(T) + AC_{ph}^{as-grown}(BT) = \gamma_n T + AC_{ph}^{as-grown}(BT)$ with the parameters $\gamma_n$, $A$, and $B$, where $\gamma_n$ is the Sommerfeld coefficient. Under the consideration of entropy balance, $\gamma_n$, $A$, and $B$ are fitted as 22.0 mJ/molK$^2$, 1.07, and 1.04 respectively. The fitting curve is shown as the dashed line in Figure S5. The value of Sommerfeld coefficient $\gamma_n$ is slightly smaller than previous reports (23 ~ 26 mJ/molK$^2$)[4,6,7,9-11], which may be originated from the vanishing residual electronic specific heat in our fully-annealed crystal. Recently, similar value of $\gamma_n$ is also reported by Xing et al.[12]. The specific heat for the as-grown crystal is larger than the fitted normal state value, which may originate from the contribution of the Schottky term from the excess Fe. After subtracting the fitted normal state specific heat data, the difference of $(C - C_n)/T$ vs $T$ for the fully-annealed crystal is shown in the right inset. Considering the electronic entropy balance, $T_c$ and the magnitude of the specific heat jump can be estimated by performing an equal area approximation for the positive and negative parts, which gives $T_c$ ~ 14.5 K, and the normalized specific heat jump at $T_c$, $\Delta C/\gamma_n T_c$, is ~ 3.0.

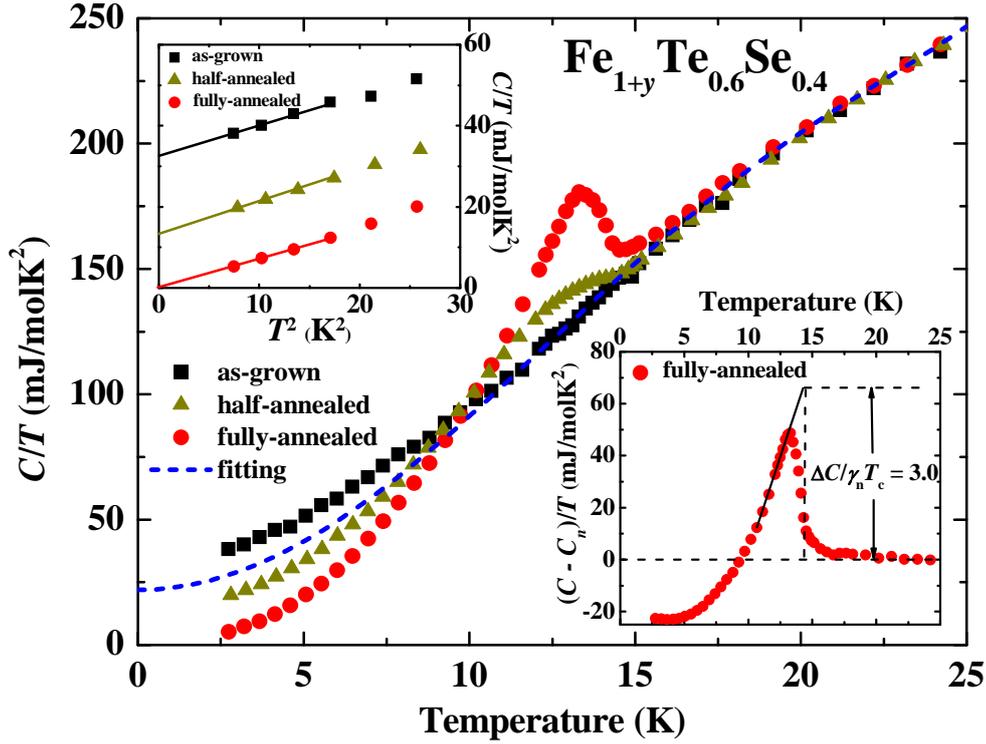

Figure S5: Temperature dependence of the specific heat plotted as $C/T$ vs $T$ for the as-grown, half- and fully-annealed $Fe_{1+y}Te_{0.6}Se_{0.4}$ single crystals. The dashed line is the fitting to the normal state specific heat of the fully-annealed crystal. The difference between the specific heat and the fitted normal state value is shown in the right inset. Left inset is the low-temperature specific heat plotted as $C/T$ vs $T^2$ for the three crystals.